# INHIBITION AND SET-SHIFTING TASKS IN CENTRAL EXECUTIVE FUNCTION OF WORKING MEMORY: AN EVENT-RELATED POTENTIAL (ERP) STUDY

Pankaj,[1] Jamuna Rajeswaran,[1*] Divya Sadana[1]

[1]Deptartment of Clinical Psychology, National Institute of Mental Health and Neurosciences (NIMHANS), Bangalore, India 560 029

### ABSTRACT

*Understanding of neuro-dynamics of a complex higher cognitive process, Working Memory (WM) is challenging. In WM, information processing occurs through four subsystems: phonological loop, visual sketch pad, memory buffer and central executive function (CEF). CEF plays a principal role in WM. In this study, our objective was to understand the neurospatial correlates of CEF during inhibition and set-shifting processes. Thirty healthy educated subjects were selected. Event-Related Potential (ERP) related to visual inhibition and set-shifting task was collected using 32 channel EEG system. Activation of those ERPs components was analyzed using amplitudes of positive and negative peaks. Experiment was controlled using certain parametric constraints to judge behavior, based on average responses in order to establish relationship between ERP and local area of brain activation and represented using standardized low resolution brain electromagnetic tomography. The average score of correct responses was higher for inhibition task (87.5%) as compared to set-shifting task (59.5%). The peak amplitude of neuronal activity for inhibition task was lower compared to set-shifting task in fronto-parieto-central regions. Hence this proposed paradigm and technique can be used to measure inhibition and set-shifting neuronal processes in understanding pathological central executive functioning in patients with neuro-psychiatric disorders.*

### KEYWORDS

*Working Memory (WM), Central Executive Function (CEF), Inhibition Task, Set-Shifting Task, Event Related Potential (ERP), sLORETA*

## 1. INTRODUCTION

Working memory (WM) is a part of the higher cognitive mental functions (HMFs) [1]. Baddeley and Hitch generated the first model of WM that represents the role of CEF [2]. WM has a temporary storage system that involves small packets of spatially distributed information processing systems [3]. It involves a four subsystem i.e., phonological loop, visual sketch pad, memory buffer and central executive function (CEF) [4]. They integrate to perform the following cognitive functions i.e., temporary storage, learning, reasoning, comprehension, manipulation and updating. This involves various parts of the sensory-motor cortices in the frontoparietal lobes, where the experience is recorded and integrated and subsequently brought to the conscious domain [5]. The information of CEF drifts through these circuits and it is coordinated by neurons in the prefrontal cortex [6]. The CEF is studied as a component of Baddeley's WM model. It's further divided in four sub components, inhibition, set-shifting, dual-tasking and updating [7].

Electroencephalography (EEG) measures the electrical activity of the brain from the electrodes on the scalp [8]. Event-related potentials (ERPs) are averaged EEG responses to a specific time-locked stimuli involving complex mental processing. ERPs are used to assess brain function and observe electro-physiological manifestations of cognitive activities in both the time domain and spatial extents. In general, ERPs are typically described in terms of components based on polarity and latency. ERPs provide good temporal resolution to certify studies of any cognitive processes in real time. When the subjects were made to perform tasks involving revision of various memories in short-term which were presented as a specific train of stimulus sequence, the responses could be measured and spatiotemporally correlated real-time using ERP [9]. EEG records the correct and incorrect responses of both the inhibition and set-shifting tasks and the ERPs associated with cognitive process and control potentials in brain are derived from this signal.

Overall, the ERPs associated with the task are correlated with CEF activity. CEF of inhibition and set-shifting tasks is manifest in the P200 (or P2), P300 (or P3) and N200 (OR N2) components of the human ERP and EEG maps show peak activity in the fronto-parietal cortices [10]. This study was aimed at understanding the brain dynamics of CEF and to find the spatial neural correlates in CEF during inhibition and set-shifting tasks. For this we designed a paradigm to study the CEF of WM and modelled the EEG signal to find out neural spatial domains during the aforementioned tasks.

# 2. MATERIAL AND METHODS

## 2.1. Subjects:

We recruited 30 educated healthy subjects (male: female=15:15, right handed, age: 28±5.9years) in NIMHANS (National Institute of Mental Health and Neuroscience, Bangalore, India) for the Event Related Potential (ERP) study. All 30 participants were selected for study after assessing them with clinical interview and normal or corrected vision (Visual Acuity < +1.5/-1.5). A written informed consent was obtained after explaining the tasks and recording procedure. Subjects with history of any neurological and mental disorders and use of substances such as alcohol and tobacco prior to session were excluded. Individuals with frequent headaches or migraine problem were excluded from the study. Completion of tasks was mandatory for inclusion in the analysis. Four participants were excluded due to bad electroencephalography (EEG) signal.

## 2.2. EEG Data Acquisitions

Electroencephalography was recorded using 32 channels Ag-AgCl electrodes arranged to the revised 10-20 system (to improve spatial sampling) and mounted in uniform fit, highly elastic cap (ECI, Electro Cap Int®, USA). EEG signal were acquired by SynAmps amplifier™ and recorded by NeuroScan™ (version 4.4) software. We used mastoid electrodes as reference. Electro-oculographic (EOG) activity was recorded by the electrodes of on the outer canthus of each eye for horizontal and below the left eye for vertical movements. Impedance of input signal of electrode was kept below less than 5kΩ (sampling rate: 1kHz). This study contains one hour EEG session. Before the start of the tasks, 3 minutes rest state EEG recording was done. Both the task paradigms were designed using "Stim2" software (Compumedics®, Neuroscan™). A button box with millisecond accuracy measured the reaction time (RT) for these computerized tasks.

## 2.3. Paradigm for the Tasks

The following two types of tasks were performed followed to eye open and eye close at relax resting condition and measure the CEF process.

#### 2.3.1 Inhibition task:

The ‘inhibition task’, subject was shown four geometric different shapes such as a triangle, circle, plus and square in four quadrants of the monitor (Figure 1). The geometric shapes were presented randomly in four quadrants checker board that were positioned in the center of the screen with a white background. The shapes are presented for 800ms at intervals of 200ms. In the four geometric shapes, if triangle (▲ ) geometric shape comes with black color, then the subject was instructed to press the button 1 or else to continue to press the button 2 of the response box. The participant was instructed to respond as quickly and accurately as possible. The whole inhibition WM task had 160 rounds and 3 minutes of completion.

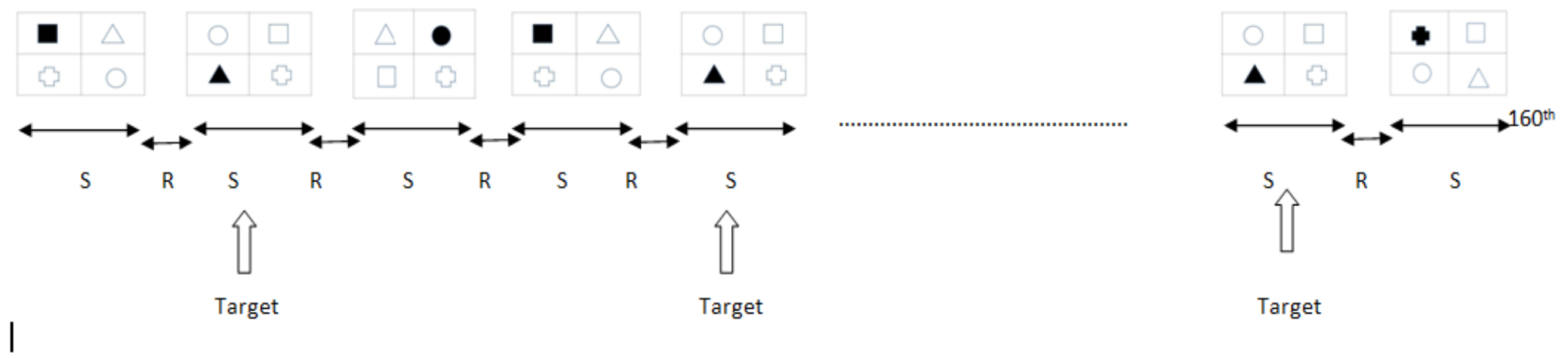

Figure 1. Experimental design for inhibition task. “S” - stimulus, which presented for 800ms, “R” - rest period kept for 200ms, “Target” – Targeted stimuli.

#### 2.3.2 Set-shifting task:

In the Set-shifting task, subjects were shown a set of patterns on a monitor (Figure 2). The set-shifting task consisted of three pictures in a row and fourth picture column was left empty. There were four options (1, 2, 3, 4) given to the subject below the pattern. Subject chooses the one of the right option. All pictures were positioned in the center of the screen in a white back ground. The subject were instructed to press the key either “1” or “2” or “3” or “4” with their response. In the task, three patterns (similar, pair, process) were present and among them each pattern had a set of 10 stimuli of similar type which were presented successively. The stimuli were presented for 4000ms at intervals of 1000ms. The participant was instructed to respond as quickly and accurately as possible. The whole set-shifting WM task had 30 rounds and 3minutes of completion.

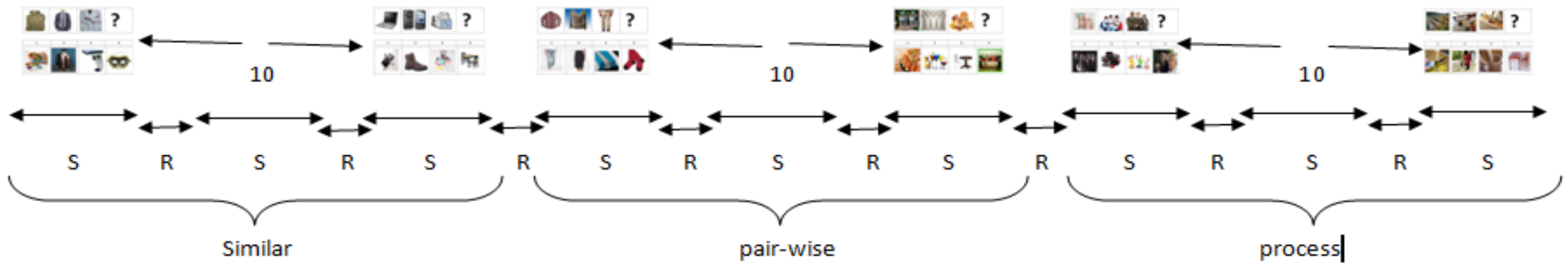

Figure 2. Experimental design for set-shifting task. “S” - stimulus, which presented for 4000ms, “R” - rest period kept for 1000ms.

## 3. DATA ANALYSIS:

### 3.1 Pre-processing and Artifact Removal from EEG signal

Raw EEG data analysis was performed on Neuroscan™ and MATLAB® software. The data was down-sampled to 250Hz. A cut-off frequency of 0.01 to 60 Hz done by 4$^{th}$ order band pass Butterworth FIR filter [9] and after 50Hz notch filter was applied for power line interference [11] [12]. Preprocessing was done by eye movement artifact correction, muscle artifact correction and singular value decomposition (SVD) for artifact free data and created linear derivation (LDR) EEG file. The artifact free EEG was epoched to a particular time window for both CEF tasks. In 'inhibition' task, EEG file were epoched to a window of 100ms pre-stimulus to 700ms post-stimulus by selecting trigger and in set-shifting task. Low pass filter of 1Hz was applied on these epoched data and were corrected to baseline. High amplitude artifacts were rejected using the artifact rejection criteria of +/-75µV. After artifact rejection, these epochs were averaged in both time and frequency domain. After this, EEG data was taken for further analysis.

## 3.2 Post Processing for EEG and ERP Signal:

ERPs were calculated from the epoched EEG data obtained by time domain averaging. Smoothening was applied on this averaged data. In the electrodes where ERP components were present as per the visual inspection, the ERP components (N200, P200, P300) were identified for inhibition task and set-shifting task. The designated time window for identification of the ERPs, which we considered in the whole experimental tasks were as follows: a) N200 - most negatively deflecting potential occurring 150ms to 250ms after the stimulus, , b) P200 - most positively deflecting potential occurring 150ms to 250ms after the stimulus and c) P300 - most positively deflecting potential occurring 250ms to 350ms after the stimulus. The group averaged EEG files contain all the ERP components to which 2D maps were generated and collected to get the activation areas during both the CEF - Working Memory tasks. We computed the activation value of amplitude in frontal cluster (F3, F7, Fz, F4, F8) and parietal cluster (P3, P7, Pz, P4, P8).

The neural activity for inhibition and set-shifting process of central executive function were mapped using "spectopo" functions of EEGLAB on 2D head view. Then the signal was computed for three dimensional cortical distributions on realistic head model with MNI152 template using standardized low resolution brain electromagnetic tomography (sLORETA). This was followed by an appropriate standardization of the current density, producing images of electric neuronal activity without localization bias [13]. sLORETA images represent the electric activity at each voxel in neuroanatomic Talairach space as the squared standardized magnitude of the estimated current density. It was found that in all noise free simulations, although the image was blurred, sLORETA had exact, zero error localization when reconstructing single sources, that is, the maximum of the current density power estimate coincided with the exact dipole location. In all noisy simulations, it had the lowest localization errors when compared with the minimum norm solution and the Dale method [14].

## 3.3 Statistics data Analysis:

Electrophysiological variables included N200, P200 and P300 peak amplitudes and latencies for Fz and Pz electrodes for both tasks. N200 to P300 latency ratio for inhibition process (Table 1) and P200 to P300 latency ratio for set-shifting process (Table 2) were used for paired T-test statistics. All ERP value found highly significant (>0.05).

Table 1. Inhibition task Paired T-test

| **Paired T-test** | | | **inhibition task** | | | |
|---|---|---|---|---|---|---|
| **frontal electrode** | stimulus | distracter | p value | t value | DF | SED |
| | P300 | P300 | 0.0001 | 17.43 | 58 | 0.34 |
| | N200 | N200 | 0.0001 | 4.83 | 58 | 0.23 |
| **parietal electrode** | stimulus | distracter | p value | t value | DF | SED |
| | P300 | P300 | 0.0001 | 9.7 | 58 | 0.91 |
| | N200 | N200 | 0.0482 | 0.76 | 58 | 0.23 |

Table 2. Set-shifting data Paired T-test

| **Paired T-test** | | | **set-shifting task** | | | |
|---|---|---|---|---|---|---|
| **frontal electrode** | value 1 | value 2 | p value | t value | DF | SED |
| | similar P300 | pair P300 | 0.2086 | 1.27 | 58 | 0.61 |
| | similar P300 | process P300 | 0.0001 | 5.18 | 58 | 0.56 |
| | pair P300 | process P300 | 0.009 | 2.7 | 58 | 0.79 |
| | similar P200 | pair P200 | 0.0004 | 3.73 | 58 | 0.18 |
| | similar P200 | process P200 | 0.9705 | 0.03 | 58 | 0.215 |
| | pair P200 | process P200 | 0.0019 | 3.25 | 58 | 0.212 |
| **parietal electrode** | value 1 | value 2 | p value | t value | DF | SED |
| | similar P300 | pair P300 | 0.0001 | 4.32 | 58 | 0.421 |
| | similar P300 | process P300 | 0.0001 | 5.69 | 58 | 0.554 |
| | pair P300 | process P300 | 0.0441 | 2.05 | 58 | 0.648 |
| | similar P200 | pair P200 | 0.0001 | 21.46 | 58 | 0.208 |
| | similar P200 | process P200 | 0.0001 | 6.73 | 58 | 0.292 |
| | pair P200 | process P200 | 0.0001 | 21.46 | 58 | 0.307 |

# 4. RESULTS:

## 4.1 Response accuracy:

The participants were performed well in CEF tasks. Accuracy and response time were analyzed in inhibition and set-shifting working memory paradigm. Comparative curves were plotted in between the inhibition and set-shifting task. We found that average score of correct response was higher in 'inhibition task' (87.5%) in comparison to 'set-shifting task' (59.5%).

## 4.2 ERP Peaks in Inhibition and Set-Shifting Task:

(a) (b)

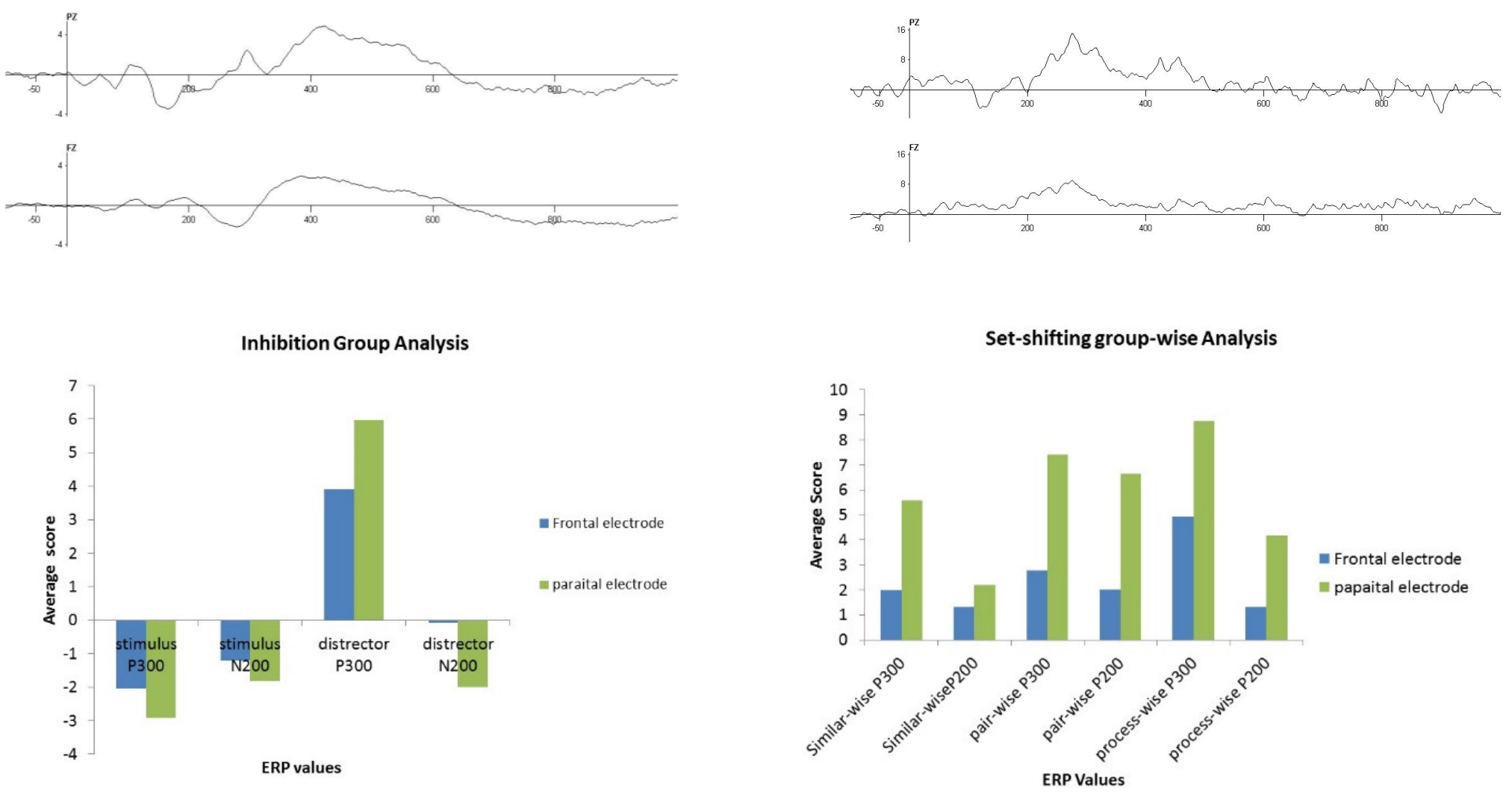

Figure 3 Event related Potentials in (a) Inhibition task (b) Set-shifting task.

Figure 3 represents the average ERPs for the N200, P200, P300 components. We measured the amplitudes for fronto-central (Fz) and parieto-central (Pz) electrode. These electrodes were significantly attenuated compared. We found that Fz electrode have less amplitude value than Pz electrode. In comparison of both tasks, inhibition process had less amplitude than set-shifting process.

## 4.3 EEG ERP results:

We found that lower peak amplitude activity of brain area in inhibition process than set-shifting process, which reveals that inhibition process utilize less activation of brain areas to perform tasks. In order to show this, we examined the frontal and parietal activity of brain in same tasks. We found that brain uses the fronto-parietal connectivity when it performs CEF of WM related tasks. We also found that the midline frontal (FZ, F4, F6) and parietal (PZ, P4, P6) electrodes were activated at the same time with prominent pattern of negative potentials and other had positive potentials in both tasks.

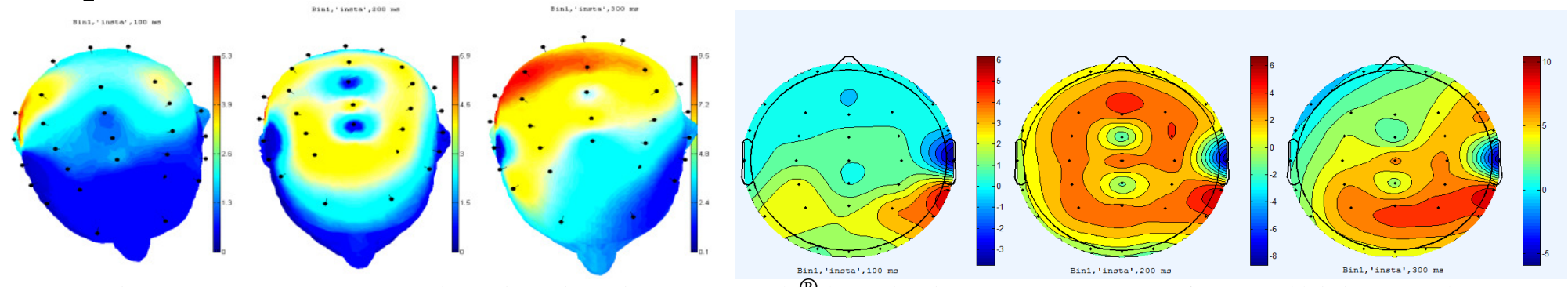

Figure 4 (a). Neuronal activation in EEGLab® head-view (3D & 2D) for Inhibition task.

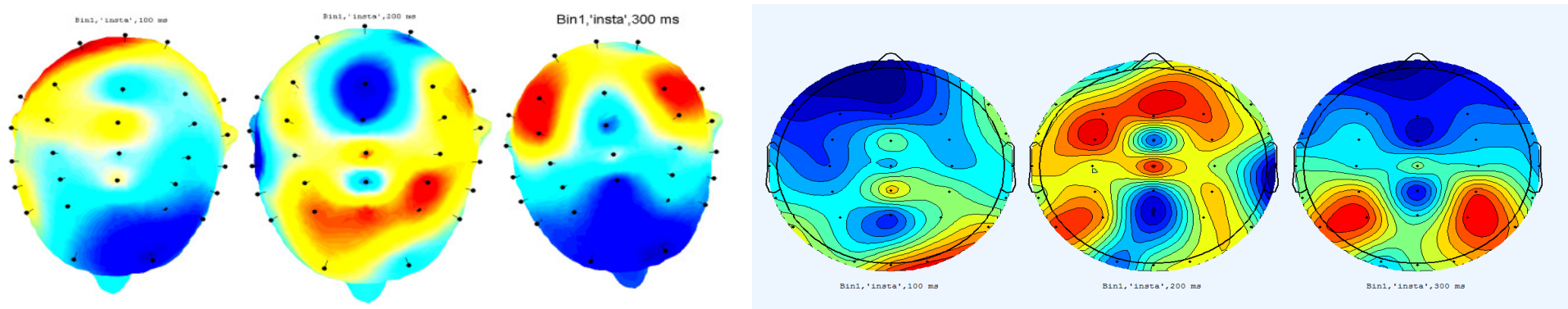

Figure 4 (b). Neuronal activation in EEGLab® head-view (3D & 2D) for Set-shifting task.

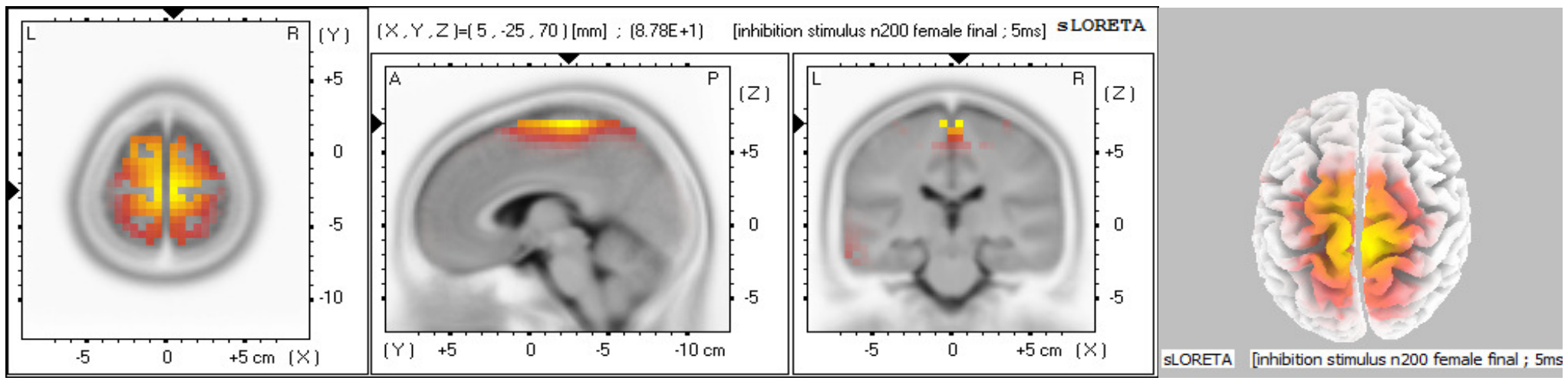

Figure 5(a) . Neuronal activation in MNI template for Inhibition task computed from EEG signal.

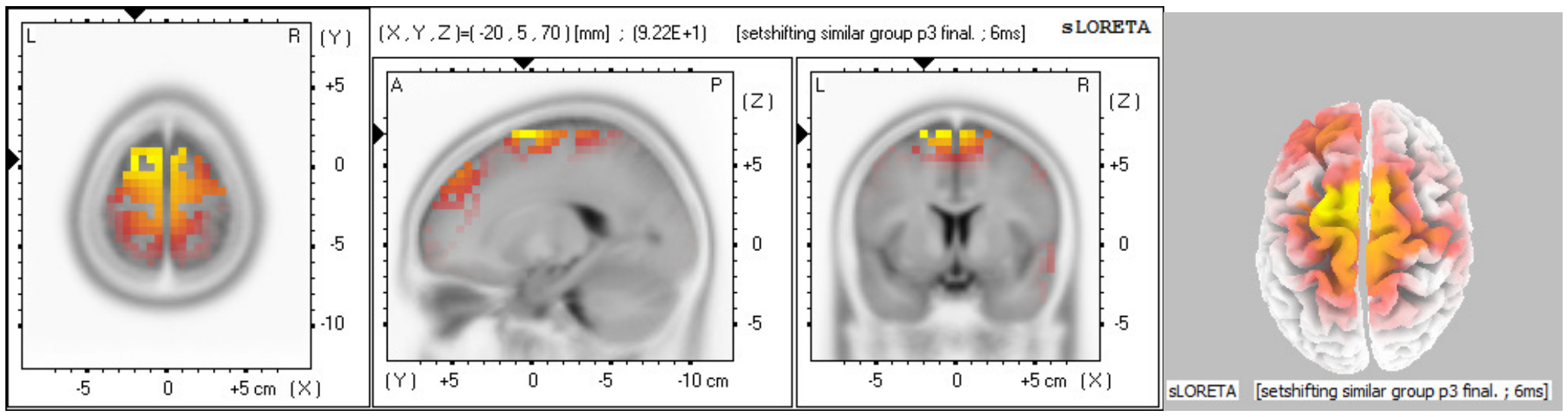

Figure 5 (b). Neuronal activation in MNI template for Set-shifting task computed from EEG signal.

Most significant results were found in both tasks. After the post processing we got 3D and 2D head-views in EEGLAB [Figure 4 (a,b)] and sLORETA [Figure 5 (a,b)]. In EEGLAB we measured the activation on 100, 200 and 300 ms after the stumuli come. In both head-views we found the Inhibition process have less activation on fronto-parietal lobs than set-shifting process.

## 5. DISCUSSION

Till today, a lot of research has been done on both central executive function of working memory but little efforts have been made to find the behavioural and neural connections between the two processes. This study examined the brain activation pattern. We find out ERPs and behavioural interpretations, during the inhibition and set-shifting tasks. In this study our objective was to understand the brain dynamics for CEF of WM functions and find out neural spatial domain in central executive function during inhibition and set-shifting process. The behavioural results showed that among the population of normal individuals, inhibition process score was high in comparison to set-shifting task.

These behavioural changes were accompanied by a number of changes in transient ERPs and task related EEG activations. This study is mainly focused on finding the EEG correlation between components of CEF. In EEG findings, by using time domain analysis [15] and bivalent 2-D and 3-D cartoon maps, positive and negative potentials were observed across the pre-frontal, frontal, central and parietal electrodes, during both the tasks.

Electric potential distribution on the basis of the scalp recording was measured by sLORETA that compute the cortical three- dimensional smoothest of all possible neural current density distribution, where neighbouring voxels have maximally similar activity. Low Resolution Electromagnetic Tomography (LORETA) is used for source localization, based on R. D. Pascual-Marqui and its virtual MR anatomical images are based on Montreal Neurological Institute (MNI) of McGill University. LORETA is a linear distribution method and it calculates the current density voxel by voxel in the brain as a linear, weighted sum of scalp electrical potential. Computations were made in a realistic head model using the MNI 152 template with the three dimensional solution space restricted to cortical grey matter.

The EEG method is the dipole source with fixed location and orientation method and it is distributed in the whole brain volume or cortical surface. There are six parameters that specify the dipole, three special coordinates (x,y,z) and three dipole components (orientation, angle, and strength). In the LORETA, the electrode potentials and matrix X are considered to be related as

$$X=LS$$

Where S is the actual current density and L is Ne×3m matrix representation the forward transmission coefficients form each source to the array of sensors. There are number of sensors Ne, extra cranial measurements in the surface of brain and Nv, voxels in the brain. In the real world applications where more concentrated focal source methods fail and choose smoothness of the inverse solution as

$$S=L^{T}(LL^{T})X$$

The sloreta uses the minimum norm solution is to create a space solution with zero contribution from the source. However, the variance of the current density estimate is based only on the measurement noise, in contrast to sLORETA , which takes into account the actual source variance as well.

**Time domain analysis of EEG and ERPs:** After averaging in time domain 2-D head maps were taken to represent the surface potential distribution patterns. Central executive WM tasks were analyzed at both the retrieval and encoding phases in epochs of 800 and 1000ms. Result was seen throughout the whole brain electrodes. In the encoding and retrieval phase of the CEF of WM activation patterns were seen [16].

**ERPs comparison studies for inhibition and set-shifting process:** Visually evoked potentials (P100) were present throughout the CEF WM task [17]. Delay in early N200 latencies in Fz of encoding showed that visual attention took time. In late P300 peaks in Pz of retrieval phase showed that matching system of visually attend objects. We also found the increase in the amplitude of the frontal visual N200 has been observed in a number of studies, this response shows the Inhibition and attention performance. These delays in ERPs are also observed in 2D and 3D maps in EEGLAB. So we can conclude that functioning of inhibition process is found in fronto-parietal area of brain.

In both encoding and retrieval phase of CEF WM all desired ERPs were found. Visually evoked potentials (P100) were seen in parietal area suggesting the stimulus representation extrastriate cortex (BA 6 and 18) [18]. N100 component was seen in frontal and pre-frontal electrodes showing the unpredictability of stimulus. N200 component was present in parietal electrodes. Encoding phase showing delayed N200 latencies but in retrieval phase N200 latencies were shorter due to attending the visual stimuli. Early-latency visual ERPs are enhanced when stimuli are presented in an attended visual field. Reaction times are also shorter to stimuli presented in an attended visual field than to those presented in an unattended visual field. In retrieval phase, delayed P200 latency of Fz electrodes suggesting for a matching system that compares sensory inputs with the stored memory. P200 latencies were present in both phases but most prominently it was elicited in retrieval phase.

ERPs analysis is suggested for activation network during CEF of WM. In Encoding phase initial activations occurred in the visual cortex where stimuli were attended for the visual search. The information from the stimuli was coded to the Fz and Pz electrodes as the late P300 peaks were present in the parietal electrodes suggesting the demand of focusing. Likewise, all the ERP patterns were seen in the retrieval phase. The only difference was the delayed P200 latencies suggesting for a matching system related to make comparison of present stimulus with the previous ones. This whole ERPs analysis is consistent with EEGLAB and sLORETA maps achieved in the time domain analysis. So, a network of activation patterns from frontal to parietal area in the both inhibition and set-shifting process of Central Executive Function of Working Memory.

## 6. CONCLUSION

Findings from this study will give a better understanding of CEF and brain activity in inhibition and set-shifting process. Our findings demonstrate a paradigm combined with a technique to be used to measure of inhibition and set-shifting neuronal processes. This would help in understanding the alteration of central executive function in patient with neurological and psychiatric disorder.

### ACKNOWLEDGMENT

Preparation of this paper was supported by Department of clinical Neuro-psychology, National Institute of Mental Health and Neuroscience (NIMHANS), Bangalore, India. We are grateful to Ganne Chaitanya, Deepak Ullal and Rajnikant Panda for providing thoughtful remarks and suggestions regarding EEG and ERP analysis.

## AUTHORS

*The Corresponding author for this paper is Dr. Jamuna Rajeswaran

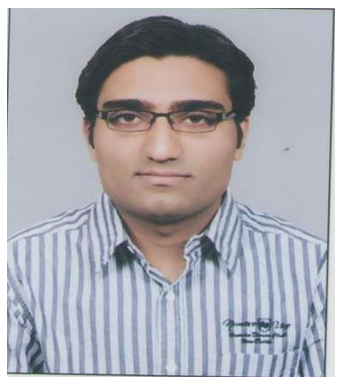

**Mr. Pankaj,** received M.Tech degree in Cognitive Neuroscience from Centre for Converging Technologies, University of Rajasthan, India, in 2015. He has more than one year of research experience on clinical psychology, Neuro-imaging and EEG signal processing on cognition at National Institute of Mental Health and Neurosciences, Bangalore, India. He has clinical research experience with many National institutes. He also got Mani Bhomik research-fellowship (2015) in National Institute of Advanced Studies (NIAS), Indian Institute of Science campus, Bangalore, India.

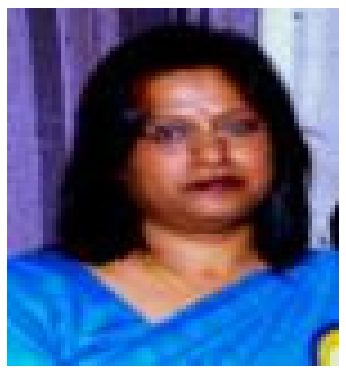

**Dr. Jamuna Rajeswaran,** Additional Professor, Consultant & Head Neuropsychology Unit, National Institute of Mental Health and Neurosciences, Bangalore, India. She completed her M.Phil and PhD from Dept. of Clinical Psychology, National Institute of Mental Health and Neurosciences, Bangalore, India.

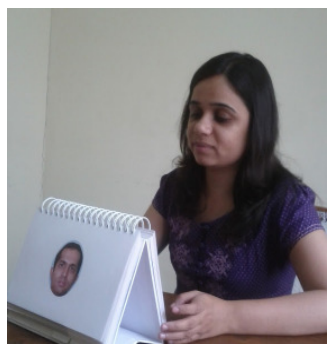

**Dr. Divya Sadana,** Senior Research Fellow, Department of Clinical psychology, National Institute of Mental Health and Neurosciences, Bangalore, India. She completed her M.Phil and PhD from Dept. of Clinical Psychology, National Institute of Mental Health and Neurosciences, Bangalore, India.